\documentclass[prl, floatfix, twocolumn, superscriptaddress]{revtex4-2}
\def\be{\begin{equation}}
\def\ee{\end{equation}}
\def\bea{\begin{eqnarray}}
\def\eea{\end{eqnarray}}
\def\bi{\begin{itemize}}
\def\ei{\end{itemize}}
\usepackage{braket}
\usepackage{xcolor}
\usepackage{graphicx}
\usepackage{amsmath,amstext,amssymb,bm,color,times}
\usepackage[english]{babel}
\usepackage[T1]{fontenc}
\usepackage[utf8]{inputenc}
\usepackage[colorlinks=true,citecolor=blue,linkcolor=magenta]{hyperref}

\begin{document}

%%%%%%%%%%%%%%%%%%%%%%%%%%%%%%%%%%%%%%%%%%%%%%%%%%%%%%%%%%%%%%%%%%%%%%%%%%

\title{Coherent Many-Body Oscillations Induced by a Superposition of Broken Symmetry States \\ in the Wake of a Quantum Phase Transition
}

%%%%%%%%%%%%%%%%%%%%%%%%%%%%%%%%%%%%%%%%%%%%%%%%%%%%%%%%%%%%%%%%%%%%%%%%%%
\author{Jacek Dziarmaga}
\affiliation{Jagiellonian University, Institute of Theoretical Physics, {\L}ojasiewicza 11, PL-30348 Krak\'ow, Poland}
\author{Marek M. Rams}
\affiliation{Jagiellonian University, Institute of Theoretical Physics, {\L}ojasiewicza 11, PL-30348 Krak\'ow, Poland}
\author{Wojciech H. Zurek}
\affiliation{Theory Division, Los Alamos National Laboratory, Los Alamos, New Mexico 87545, USA}

\date{January 29, 2022}
%%%%%%%%%%%%%%%%%%%%%%%%%%%%%%%%%%%%%%%%%%%%%%%%%%%%%%%%%%%%%%%%%%%%%%%%%%

\begin{abstract}
It is now widely accepted that quenches through the critical region of quantum phase transitions result in post-transition states populated with topological defects -- analogs of the classical topological defects. However, consequences of the very non-classical fact that the state after a quench is a {\it superposition} of distinct, broken--symmetry vacua with different numbers and locations of defects have remained largely unexplored. We identify coherent quantum oscillations induced by such superpositions in observables complementary to the one involved in symmetry breaking. These oscillations satisfy Kibble-Zurek dynamical scaling laws with the quench rate, with an instantaneous oscillation frequency set primarily by the gap of the system. In addition to the obvious fundamental significance of a superposition of different broken symmetry states, quantum coherent oscillations can be used to verify unitarity and test for imperfections of the experimental implementations of quantum simulators.
\end{abstract}

\maketitle

%%%%%%%%%%%%%%%%%%%%%%%%%%%%%%%%%%%%%%%%%%%%%%%%%%%%%%%%%%%%%%%%%%%%%%%%%%%%%%

%%%%%%%%%%%%%%%%%%%%%%%%%%%%%%%%%%%%%%%%%%%%%%%%%%%%%%%%%%%%%%%%%%%%%%%%%%%%%%

{\it Motivation.---}%
Studies of quenches 
%that take the system 
through a symmetry-breaking quantum phase transition at a finite rate have been to a large extent focused on the generation of topological defects. This was clearly the first thing to do, as topological defects are stable and the obvious focus of interest in the classical (i.e., thermodynamic) nonequilibrium phase transitions. 
By contrast, quantum phase transitions inevitably lead to superpositions of the eigenstates of the post–transition Hamiltonian. Such superpositions 
%of energy eigenstates 
(in, e.g., an atom) result in oscillations with the frequency given by the difference between the energies of the two levels involved, and 
%with 
the amplitude set by their initial occupancy. We show that 
%such 
superpositions of the post-transition eigenstates are inevitable in quantum phase transitions and exhibit analogous (many-body) coherent quantum oscillations. We characterize their appearance and properties 
%as a function of the quench rate 
in models where they can be investigated analytically or numerically, and where they should be accessible to experiments. 

The obvious motivation for investigating collective oscillations of many-body systems is because they are there, and because they are a signature of the quantumness of the transition. Moreover, such an oscillatory behavior 
%also 
constitutes a sensitive probe of the imperfections of the experiment, including especially decoherence. We show that the form of the oscillations is simple when the energy levels of the many-body system are degenerate (as then the number of frequencies involved is small). When the degeneracies are lifted by the imperfections of the Hamiltonian (e.g., caused by its implementation), dephasing will result in the loss of coherence. Furthermore, decoherence caused by imperfect isolation of the system 
%from other (external or internal) degrees of freedom 
will result in non-unitary evolution causing a further gradual loss of coherence.
Therefore, such oscillations can serve as a diagnostic tool to assess how accurate -- and especially how quantum -- is the implementation of the transition in the emulation experiments: There are now examples of quantum phase transitions that are both solvable and experimentally accessible, creating appealing possibilities to use the exact many-body time-dependent solutions to benchmark experimental implementations. The post-transition oscillations should be relatively easy to prepare and detect in contrast to the more challenging non-local ``double slit - like'' superpositions of topological defects~\cite{cats_nature_phys}.

%%%%%%%%%%%%%%%%%%%%%%%%%%%%%%%%%%%%%%%%%%%%%%%%%%%%%%%%%%%%%%%%%%%%%%%%%%%%%%

{\it Kibble-Zurek mechanism.---}%
The Kibble-Zurek mechanism (KZM) has its roots in 
%a scenario for defect creation in 
cosmological symmetry-breaking phase transitions~\cite{K-a, *K-b, *K-c}. Kibble considered cooling Universe where causally disconnected regions 
%must 
independently select broken symmetry vacua. This mosaic of broken symmetry domains leads to topologically nontrivial configurations.
%where t
The extent of such
% of correlated 
domains is limited by the 
%finite 
size of the causal horizon. 

This cosmological constraint is not relevant for laboratory experiments. Therefore, a dynamical theory for the continuous phase transitions 
was proposed and developed~\cite{Z-a, *Z-b, *Z-c, Z-d}. KZM employs equilibrium critical exponents to predict the scaling of the defects density as a function of the quench rate. It has been verified in numerous simulations~\cite{KZnum-a,KZnum-b,KZnum-c,KZnum-d,KZnum-e,KZnum-f,KZnum-g,*KZnum-h,*KZnum-i,KZnum-j,KZnum-k,KZnum-l,KZnum-m,that} and condensed matter experiments~\cite{KZexp-a,KZexp-b,KZexp-c,KZexp-d,KZexp-e,KZexp-f,KZexp-g,KZexp-gg,KZexp-h,KZexp-i,KZexp-j,KZexp-k,KZexp-l,KZexp-m,KZexp-n,KZexp-o,KZexp-p,KZexp-q,KZexp-r,KZexp-s,KZexp-t,KZexp-u,KZexp-v,KZexp-w,KZexp-x}. Topological defects are central in those studies, as they can persist despite dissipation 
%that is 
inevitable in thermodynamic systems.

The quantum version of KZM (QKZM) considers quenches across quantum critical points. It has been developed~\cite{QKZ1,QKZ2,QKZ3,d2005,d2010-a, d2010-b, QKZteor-a,QKZteor-b,QKZteor-c,QKZteor-d,QKZteor-e,QKZteor-f,QKZteor-g,QKZteor-h,QKZteor-i,QKZteor-j,QKZteor-k,QKZteor-l,QKZteor-m,QKZteor-n,QKZteor-o,KZLR1,KZLR2,QKZteor-oo,delcampostatistics,KZLR3,QKZteor-q,QKZteor-r,QKZteor-s,QKZteor-t,sonic,QKZteor-u,QKZteor-v,QKZteor-w,QKZteor-x, roychowdhury2020dynamics,sonic,schmitt2021quantum} and put to experimental tests~\cite{QKZexp-a, QKZexp-b, QKZexp-c, QKZexp-d, QKZexp-e, QKZexp-f, QKZexp-g,deMarco2,Lukin18,adolfodwave,2dkzdwave}. Recent experiments 
%are 
target the exactly solvable quantum Ising chain in the transverse field, employing simulators based on Rydberg atoms~\cite{Lukin18} and superconducting qubits~\cite{King_Dwave1d_2022}. Scaling of the resulting defects densities appears to be consistent with the 
%analytical solution quantifying defect generation
QKZM predictions~\cite{QKZ2,QKZ3,d2005}. Ongoing experimental developments~\cite{rydberg2d1,rydberg2d2,Semeghini2021,Satzinger2021etal} open possibility to study the quantum dynamics in two-dimensional systems. 

%Notwithstanding such impressive advances, we note that 
Of course, by the time defects are counted, quantum superpositions that should be present in the post-transition state are long gone. Thus, the {\it quantumness} of phase transition dynamics has not been, as yet, certified in the experiments. Indeed -- as approximate scalings observed are not a unique fingerprint of the defect formation mechanism, and is not clear at what stage the systems used in the experiments decohere and become effectively classical -- it would be desirable to directly verify quantumness of the phase transition dynamics. Coherent oscillations we are describing offer that possibility.
% to do so. 
They can also be used to benchmark quantumness of the hardware used in (e.g., adiabatic) quantum computing. 

%%%%%%%%%%%%%%%%%%%%%%%%%%%%%%%%%%%%%%%%%%%%%%%%%%%%%%%%%%%%%%%%%%%%%%%%%%%%%%

A smooth ramp crossing the critical point at time $t_c$ can be linearized in its vicinity as 
\begin{equation}
\epsilon(t)=\frac{t-t_c}{\tau_Q},
\label{epsilont}
\end{equation}
where $\epsilon$ measures the distance from the quantum critical point and quench rate is given by $\tau_Q$. The system is prepared in the ground state far from the critical point. The initial evolution adiabatically follows the time-dependent Hamiltonian. This approximate adiabaticity fails at time $-\hat t$ 
%(measured from 
before $t_c$ when the reaction rate of the system (set by the gap) becomes comparable to the instantaneous relative ramp rate, namely $\Delta\propto|\epsilon|^{z\nu} \propto |\dot \epsilon/\epsilon| = 1/|t|$. 
This leads to characteristic timescale 
\be 
\hat t\propto \tau_Q^{z\nu/(1+z\nu)},
\label{hatt}
\ee
where $z$ is dynamical critical exponent, and $\nu$ is correlation length exponent~\cite{Z-a}. In the adiabatic-impulse-adiabatic scenario, the ground state at $-\hat\epsilon = -\hat t/\tau_Q\propto -\tau_Q^{-1/(1+z\nu)}$ fluctuating on a scale set at  $-\hat t$ survives until $+\hat t$, and the correlation length, 
\begin{equation}
\hat\xi \propto \tau_Q^{\nu/(1+z\nu)},  
\label{hatxi}
\end{equation}
becomes imprinted for the subsequent adiabatic evolution. This oversimplified scenario correctly predicts the scaling dependence of the characteristic length and time scales on $\tau_Q$. They naturally appear in KZM dynamical scaling hypothesis~\cite{KZscaling1,KZscaling2,Francuzetal}. For an observable ${\cal O}$,
\be 
\hat\xi^{\Delta_{\cal O}} \bra{\psi(t)} {\cal O}_r \ket{\psi(t)} = F_{\cal O}\left((t-t_c)/\hat\xi^z,r/\hat\xi\right),
\label{KZscalingO}
\ee
where $\ket{\psi(t)}$ is the state of the system, $\Delta_{\cal O}$ is the scaling dimension, $F_{\cal O}$ is a non-universal scaling function, and $r$ is a distance in, e.g., a correlation function. It is expected to hold in the vicinity of the critical point, for $t$ between 
%roughly 
$\pm\hat t$.

%%%%%%%%%%%%%%%%%%%%%%%%%%%%%%%%%%%%%%%%%%%%%%%%%%%%%%%%%%%%%%%%%%%%%%%%%%%%%%

In the following, we employ the paradigmatic Ising Hamiltonian in a transverse field, \begin{equation}
H(t) = - J(t) \sum_{\langle m,n\rangle} \sigma^z_m \sigma^z_n - g(t) \sum_m \sigma^x_m.
\label{eq:hamiltonian}
\end{equation}
Here, $\sigma_m^{x}$, $\sigma_m^{y}$, and $\sigma_m^{z}$ denote the Pauli matrices on lattice site $m$, and 
%we will consider 
interactions that are between neighboring sites, $\langle m,n\rangle$. We consider three lattice geometries: (i) an integrable one-dimensional chain (1D) where each site has two neighbors, and two non-integrable models where each site has four neighbors: (ii) a 1D ladder where sites that are next-nearest neighbors in a chain become adjacent, and (iii) a two-dimensional square lattice geometry (2D). We pictorially represent those lattice geometries as insets in Figures.

%%%%%%%%%%%%%%%%%%%%%%%%%%%%%%%%%%%%%%%%%%%%%%%%%%%%%%%%%%%%%%%%%%%%%%%%%%%%%%

%%%%%%%%%%%%%%%%%%%%%%%%%%%%%%%%%%%%%%%%%%%%%%%%%%%
\begin{figure}[t!]
\begin{centering}
\includegraphics[width=\columnwidth]{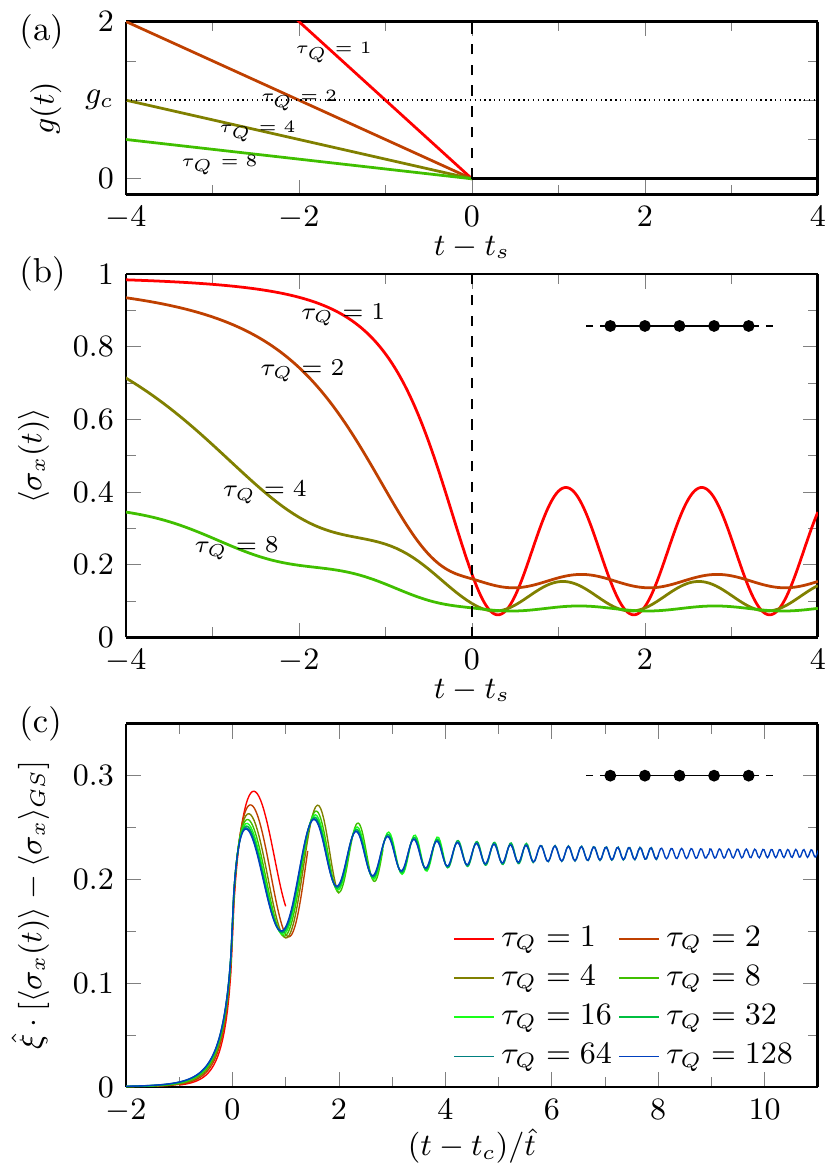}
\par\end{centering}
\caption{
{\bf Coherent oscillations after a quench through a quantum critical point.} 
In panel (a), we show a quench protocol in the 1D transverse-field Ising model, where we allow the system to freely evolve upon reaching zero transverse field at time $t=t_s$. 
In (b), we track transverse magnetization in the 1D model, where the coherent oscillations for $t>t_s$ are apparent.
Their origin can be traced back to the point reached at $t_c$, see panel (c), where we show data collapse consistent with the dynamical scaling hypothesis in Eq.~\eqref{Xscaling}. There is a gradual decrease of oscillation amplitude
during the ramp through the ferromagnetic phase, setting the nonzero amplitude observed for~$t>t_s$ (note that panel (c) shows only $t<t_s$).}
\label{fig:oscillations_1D}
\end{figure}
%%%%%%%%%%%%%%%%%%%%%%%%%%%%%%%%%%%%%%%%%%%%%%%%%%%

{\it Oscillations in 1D.---}%
We begin with the 1D version~\cite{QKZ2,d2005,QKZteor-d,Francuzetal,RadekNowak} where we traditionally set $J=1$ and 
%linearly 
ramp the transverse field,
\begin{eqnarray}
    g(t) = g_c (1 - \epsilon(t)) = g_c - g_c (t-t_c) / \tau_Q,
    \label{ramp}
\end{eqnarray}
from $t=-\infty$ in the limit of strong 
%transverse 
field, across the critical point at $g(t_c) = g_c=1$, to $g(t_s)=0$ where the transverse field vanishes. The Jordan-Wigner transformation maps the model to a set of independent two-level Landau-Zener systems that can be solved analytically. In particular, the final density of excited quasiparticles/kinks scales like~\cite{QKZ2,d2005,SM}
\be
n\approx\frac{1}{2\pi\sqrt{2\tau_Q}} \propto \hat\xi^{-1},
\label{n}
\ee
consistent with the critical exponents $z=\nu=1$. The average defect density (accessed by counting them in the experiments to date) is a very superficial characterization of the final state, which, in fact, should be -- prior to the kink count -- a quantum superposition of different numbers~\cite{QKZteor-d,delcampostatistics} and correlated locations of kinks~\cite{roychowdhury2020dynamics,RadekNowak}.

Breaking 
%away 
with tradition, we do not focus on kinks but rather on the transverse magnetization, $\sigma^x$, that does not commute with the kink observables. Its expectation value during and after crossing the critical point is shown in Fig.~\ref{fig:oscillations_1D}, where we consider linear ramps~\eqref{ramp} with several values of the quench time. All the ramps stop at $g=0$, allowing the system to freely evolve with a purely ferromagnetic Hamiltonian for $t>t_s$. In accordance with the general QKZM scaling hypothesis~\eqref{KZscalingO}, for slow enough $\tau_Q$ the transverse magnetization in the vicinity of the critical point should satisfy
\begin{equation}
    \hat\xi^{\Delta_x} 
    \left[  
    \langle \sigma^x(t) \rangle - \langle \sigma^x \rangle_{\rm GS}
    \right]=
    F_{\sigma^x}\left[(t-t_c)/\hat t\right].
    \label{Xscaling}
\end{equation}
Here $\langle \sigma^x \rangle_{\rm GS}$ is transverse magnetization in the instantaneous ground state for transverse field $g(t)$, and the scaling dimension $\Delta_x=1$ for a 1D chain. As we can see in Fig.~\ref{fig:oscillations_1D}(c), the KZM-rescaled plots for different quench timescales $\tau_Q$ collapse to a common scaling function.  In this integrable case good collapse extends beyond $+\hat t$. The function is oscillatory with an instantaneous frequency dominated by twice the quasiparticle gap as two quasiparticles with opposite quasimomenta are the relevant excitation. The amplitude of the oscillations slowly decays with the scaled time, partly due to a dephasing by a non-trivial quasiparticle dispersion and partly due to the adiabatic evolution of the excited Bogoliubov modes. 

%%%%%%%%%%%%%%%%%%%%%%%%%%%%%%%%%%%%%%%%%%%%%%%%%%%
\begin{figure}[t!]
\begin{centering}
\includegraphics[width=\columnwidth]{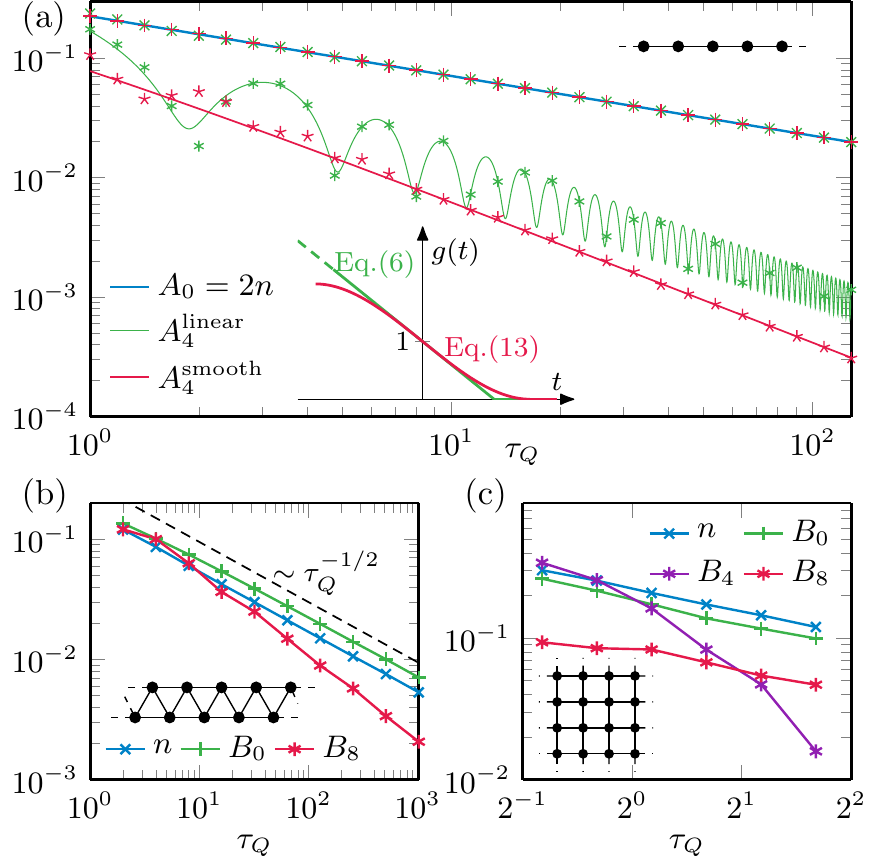}
\par\end{centering}
\caption{{\bf Scaling of transverse magnetization oscillations amplitudes}. 
In (a), for a 1D chain, we show constant contribution, $A_0$ (blue line), and the amplitudes of coherent oscillations, $A_4$, as a function of 
%the quench rate 
$\tau_Q$. We compare the 
%leading 
analytical formulas in Eqs.~\eqref{eq:1d_Xt} and~\eqref{eq:1d_Xt_smooth} (lines) with the corresponding exact numerical results (points). We show the results for two quench protocols of the same linear slope $\propto \tau_Q^{-1}$ at the critical point: the first protocol in Eq.~\eqref{ramp} stops abruptly at $g=0$ (green line indicates the amplitude of oscillations), and the second one in Eq.~\eqref{ramp_smooth} reaches $g=0$ smoothly (red). The protocols are shown in the inset.
In (b), for the ladder, the constant term $B_0$ closely follows the measured density of excitations, $n$. The latter follows an expected scaling for a model in the same universality class as a 1D chain, $n\sim \tau_Q^{-1/2}$. The dominant oscillatory contribution, $B_8$, can be fitted with $B_8~\sim \tau_Q^{-0.69}$, though we expect that logarithmic corrections from dephasing contribute to the decay of $B_8$. Note that in (a) the red line would be consistent with a power-law and exponent $-1.18$ while the analytical solution shows $n^2 \sim\tau_Q^{-1}$ behavior times a logarithmic correction. 
In (c), we show the data for the 2D square lattice where the dominant contributions, $B_0$ and $B_8$, closely follow the measured excitation density for available times.
}
\label{fig:1D_oscillations}
\end{figure}
%%%%%%%%%%%%%%%%%%%%%%%%%%%%%%%%%%%%%%%%%%%%%%%%%%%

The ramps in Fig.~\ref{fig:oscillations_1D}(a) terminate at $g=0$ where the transverse magnetization in the ground state is zero: $\langle \sigma^x \rangle_{\rm GS}=0$. Therefore, $F(\infty) \hat\xi^{-1}\propto \tau_Q^{-1/2}$ is the initial transverse magnetization for the subsequent free evolution with $g=0$, where
\begin{eqnarray}
     \langle 
     \sigma^x_m(\tau)
     \rangle =
     \langle e^{i \tau H} \sigma^x_m e^{-i \tau H} \rangle_{t_s} = 
 A_0 + A_{4} \cos(4 \tau + \tilde \phi), ~ ~
\end{eqnarray}
as each site is uniformly coupled to 2 neighbours. Here, we introduce 
$$\tau = t - t_s > 0,$$
as the duration of free evolution with $g=0$. As we can see there is a constant term plus oscillations with a single frequency. The amplitudes are determined from expectation values in the state at $t=t_s$ at the end of the linear ramp and the beginning of the free evolution~\cite{SM}. 

From the exact solution~\cite{SM}, we can extract an asymptotic form for $\tau_Q\gg 1$:
\begin{equation}
    \label{eq:1d_Xt}
    \langle 
    \sigma^x_m(\tau)
    \rangle
    = 2n + 2C d n^2 \cos(4\tau+\phi)-\pi^2n^2\sin(4\tau).
\end{equation}
Here, $A_0= 2n$ is set by the density of kinks in Eq.~\eqref{n}, which is conserved for $t>t_s$. The amplitude
\begin{equation}
    \label{eq:1d_amplitude}
    A_4^{\rm{linear}}= n^2 \sqrt{\pi^4+4\pi^2 C d \sin\phi +4C^2d^2},
\end{equation}
where $C\approx 57\sqrt{6\pi}/80$ is a numerical constant, $\phi$ is a phase accumulated by the KZ-excited quasiparticles~\cite{SM}, and 
\begin{equation}
d =
\left[ 
1 +
\left( 
{3\ln\tau_Q} / {4\pi} 
\right)^2
\right]^{-3/4} < 1
\label{d_linear}
\end{equation}
is a factor due to dephasing of the KZ excitations by their non-trivial dispersion. The constant term and the amplitude are plotted in Fig.~\ref{fig:1D_oscillations}(a) as functions of the quench time $\tau_Q$.

As we can see, the amplitude is not a simple power-law in $\tau_Q$. Irregularities originates from interference between the two oscillatory contributions to~\eqref{eq:1d_Xt}, from the KZ excitation near the critical point, $\propto \cos(4\tau+\phi)$, and from the abrupt termination of the linear ramp at $g=0$, $\propto \sin(4\tau)$.  To focus on  KZ oscillations we eliminate the non-KZ oscillations~\footnote{To be more precise, making them higher order in powers of $n$} by using, instead of the all-linear ramp in Eq.~\eqref{ramp}, a smoother version,
\begin{equation}
    \tilde \epsilon(t)= \frac{t-t_c}{\tau_Q} - \frac{4}{27} \left(\frac{t-t_c}{\tau_Q}\right)^3,
    \label{ramp_smooth}
\end{equation}
replacing $\epsilon(t)$ in Eq.~\eqref{ramp}. This protocol starts in the ground state at $g(t_c - \frac{3}{2}\tau_Q) = 2$, and terminates at $g(t_s)=0$, for $t_s=t_c+\frac{3}{2}\tau_Q$, with a zero time derivative, $\dot g(t_s)=0$. This leads to pure post-KZ oscillation amplitude,
\begin{equation}
A_4^{\rm{smooth}} = 2C d ~n^2,
\label{eq:1d_Xt_smooth}
\end{equation}
that scales simply as $n^2\propto\tau_Q^{-1}$, with a logarithmic correction brought in by the dephasing factor (see, Fig.~\ref{fig:1D_oscillations}). The latter is slightly reduced, replacing $\ln \tau_Q$ with $0.2164 + \ln \tau_Q$ in Eq.~\eqref{d_linear}, as the approach to $g=0$ makes the smooth ramp longer. However, the reduction is negligible when $0.2164\ll\ln\tau_Q$, because the extra time needed for the smooth ending of the ramp is spent mostly near $g=0$, where the quasiparticle dispersion is almost flat, and there is little extra dephasing.

The smooth ramp is not the only way to eliminate non-KZ oscillations. For instance, an imperfect termination of the linear ramp at a finite $g_f\ll1$ (instead of $0$) results in  a gradual suppression of the oscillations with time. The small finite transverse field means that the quasiparticle dispersion is non-trivial although almost flat. The non-KZM excitations, that span all quasi-momenta, dephase after time $\propto 1/g_f$. On the other hand, the influence on the KZ-part appears in the dephasing factor, replacing $\ln \tau_Q$ with $\ln \tau_Q - g_f^2 + 2 \tau g_f / \tau_Q$ in Eq.~\eqref{d_linear}.  The KZ-part that originates from small quasi-momenta modes, becomes suppressed when $\tau\gg\tau_Q/g_f$. For large enough $\tau_Q$, it becomes much larger than the dephasing time of the non-KZ part, thus opening a time window when the non-KZ oscillations are suppressed but the KZ ones are not.  It highlights the stability of KZ-related oscillations.

%%%%%%%%%%%%%%%%%%%%%%%%%%%%%%%%%%%%%%%%%%%%%%%%%%%%%%%%%%
\begin{figure}[t!]
\begin{centering}
\includegraphics[width=\columnwidth]{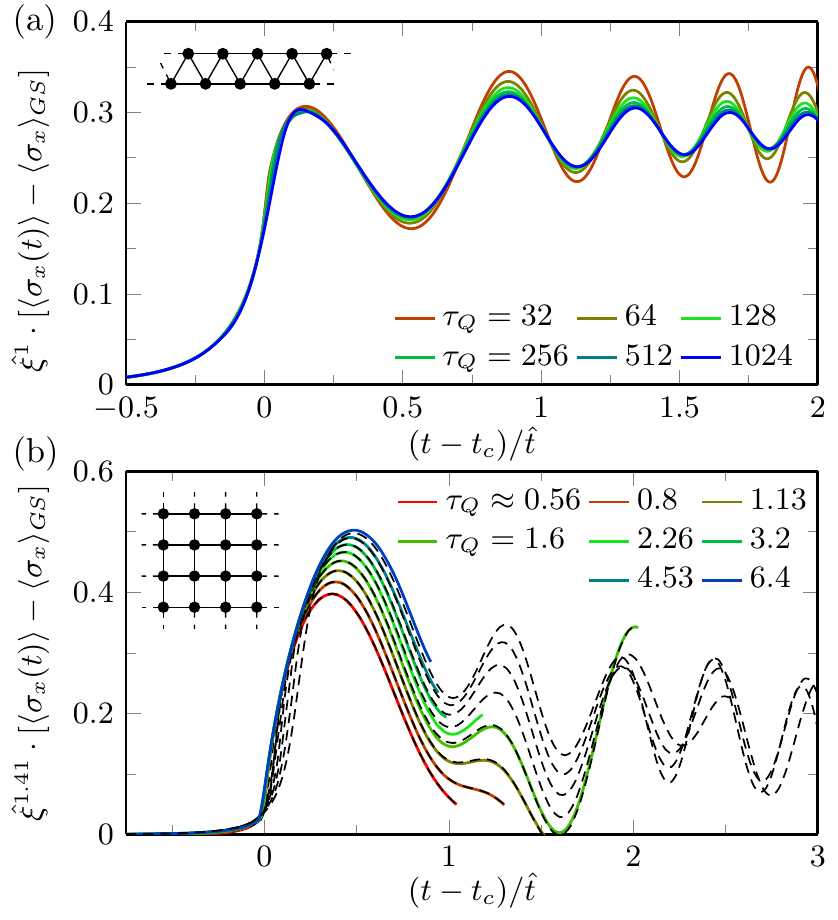}
\par\end{centering}
\caption{
{\bf Oscillations in non-integrable systems. }
In (a), we show scaled transverse magnetization~\eqref{Xscaling} in the function of scaled time in the 1D ladder geometry. The critical point of the model is in the same universality class as a 1D chain, with $\nu=z=\Delta_x =1$. For large enough $\tau_Q$, the plots collapse in the vicinity of the critical point to a single scaling function $F_{\sigma^x}$ that exhibits oscillatory behavior. 
In (b), the corresponding data for the 2D transverse Ising model with the scaling dimension $\Delta_x \simeq 1.41$.
The values of $\tau_Q$ are limited, from below, to be in the scaling limit at the critical point and, from above, to avoid finite-size effects in a finite lattice~\cite{schmitt2021quantum}. The best collapse of complementary quantities used $\xi = \tau_Q^{0.36}$~\cite{schmitt2021quantum} for similar range of $\tau_Q$'s.
Solid lines indicate thermodynamic limit results of iPEPS~\cite{gradient}, which become unstable for times longer than shown. Dashed lines indicate the MPS results measured in the center of a finite $11 \times 11$ lattice.
}
\label{fig:oscillations2D}
\vspace{-5pt}
\end{figure}
%%%%%%%%%%%%%%%%%%%%%%%%%%%%%%%%%%%%%%%%%%%%%%%%%%%%%%%%%%
%

{\it Oscillations in non-integrable systems.---}%
Qualitatively similar results can be obtained for non-integrable systems though they make us resort to numerical simulations, see Fig.~\ref{fig:oscillations2D}. For a 1D ladder, we use uniform matrix product states (uMPS)~\cite{Vanderstraeten2019} for a system in thermodynamic limit, and in 2D either the MPS~\cite{Haegeman2016} on a $11\times11$ lattice or the iPEPS in the thermodynamic limit~\cite{gradient}.
We employ a protocol that is gradually turning on the Ising terms while turning off the transverse field~\footnote{We considered a similar protocol for a 2D system in Ref.~\cite{schmitt2021quantum}. Comparing to that work, here, we rescale the Hamiltonian (and all time-scales) by a factor of $2$, to have $J=1$ when $g=0$.},
\begin{eqnarray}
g(t)/g_c = {(1 - \epsilon(t))}/{2},~~~ J(t) = {(1 + \epsilon(t))}/{2}.
\label{eq:ramp2D} 
\end{eqnarray}
We use the linear ramp in Eq.~\eqref{epsilont} for the ladder and a smooth ramp in Eq.~\eqref{ramp_smooth} for the square lattice.
The models exhibit phase transitions (for $J=1$) at $g_c\approx3.04438$ in 2D~\cite{Deng_QIshc_02} and we identify $g_c\approx2.4785$ for the ferromagnetic ladder.
 
After the ramp ends, at $g=0$, the oscillations continue as
\begin{equation}
    \langle  \sigma^x_m(\tau) \rangle =  B_0 + B_{4} \cos(4 \tau+\phi_{4}) + B_{8} \cos(8 \tau+\phi_{8}), 
\end{equation} 
with a constant term and two frequencies of oscillations, resulting from an uniform Ising coupling of a site to 4 neighbouring sites~\cite{SM}. The amplitudes are shown in Fig. \ref{fig:1D_oscillations}.

Fig.~\ref{fig:oscillations2D} is testing the scaling hypothesis~\eqref{Xscaling} for the non-integrable models.
With increasing $\tau_Q$ the plots tend to an oscillatory scaling function in the vicinity of the critical point
even though in 2D, due to the growth of entanglement with increasing $\tau_Q$, our simulations are limited to relatively fast transitions (i.e., quench times where the integrable 1D Ising also exhibits discrepancies with the limiting slow quench behavior).

{\it Conclusion.---}%
%%%%%%%%%%%%%%%%%%%%%%%%%%%%%%%%%%%%%%%%%%%%%%%%%%%%%%%%%%%%%%%%%%%%%%%%%%%%%%
The post-quench state is a superposition of different numbers of kinks (excited bonds). 
Two manifolds of eigenstates that differ by $m$ excited bonds ($m=2$ for a chain, and  $m=4$ for a ladder) result in oscillations:
\be 
\ket{...\downarrow\downarrow\downarrow\downarrow\downarrow...}+
{\cal A}
\ket{...\downarrow\downarrow\uparrow\downarrow\downarrow...}    
e^{-2m it}.
\label{flip}
\ee 
This is the most obvious quantum signature of the consequences of the quantum phase transition. 

For a chain the probability of a single-spin flip is $|{\cal A}|^2\propto n^4$ (with a logarithmic correction) in agreement with~\cite{RadekNowak,dziarmaga_kinks_2022} where antibunching of kinks makes it decay faster than $n^2$.
For a ladder we fit $|{\cal A}|^2 \propto\tau_Q^{-1.1}$. 
In both cases the amplitude of the oscillations follows as a square root of the probability.
In 2D, the higher oscillation frequency, $8$, similarly comes from isolated spin flips similar to~\eqref{flip}, while the lower frequency, $4$, and the constant term are due to spin flips adjacent to (coarse) domain walls.
In 1D cases the dominant energy eigenvalues have nearly the same separation (a multiple of the gap) so the oscillation occurs with an essentially single (time dependent) frequency. In the 2D case the picture becomes slightly more complicated, but the few frequencies are still controlled by the gap size. 

The secular part of the response to the quench follows from the same treatment and is also quantum, but the oscillatory part is a more compelling signature of the quantumness of the transition. Coherent oscillations are vulnerable to decoherence (see~\cite{cats_nature_phys} for related discussion) and to imperfect implementation of the Hamiltonian. Decoherence that einselects broken symmetry states is plausible in many-body systems. It will localize kinks while suppressing oscillations, as do the measurements aimed at testing KZM performed to date. Pointer observable is einselected at least in part by the system-environment coupling~\cite{ZRMP}, so e.g. ``quantum limit of decoherence'' that favors energy eigenstates, \cite{PZ99} is also possible.

%%%%%%%%%%%%%%%%%%%%%%%%%%%%%%%%%%%%%%%%%%%%%%%%%%%%%%%%%%%%%%%%%%%%%%%%%%%%% 
%\acknowledgements
%%%%%%%%%%%%%%%%%%%%%%%%%%%%%%%%%%%%%%%%%%%%%%%%%%%%%%%%%%%%%%%%%%%%%%%%%%%%%
%
This research was 
funded by the National Science Centre (NCN), Poland, under project 2021/03/Y/ST2/00184 within the QuantERA II Programme that has received funding from the European Union’s Horizon 2020 research and innovation programme under Grant Agreement No 101017733 (JD),
NCN under project 2020/38/E/ST3/00150 (MMR), and
Department of Energy under the Los Alamos National Laboratory LDRD Program (WHZ).
WHZ was also supported by the U.S. Department of Energy, Office of Science, Basic Energy Sciences, Materials Sciences and Engineering Division, Condensed Matter Theory Program.
%
%%%%%%%%%%%%%%%%%%%%%%%%%%%%%%%%%%%%%%%%%%%%%%%%%%%%%%%%%%%%%%%%%%%%%%%%%%%%

%%%%%%%%%%%%%%%%%%%%%%%%%%%%%%%%%%%%%%%%%%%%%%%%%%%%%%%%%%%%%%%%%%%%%%%%%%%%
\bibliographystyle{apsrev4-2}
\bibliography{KZref.bib}
%%%%%%%%%%%%%%%%%%%%%%%%%%%%%%%%%%%%%%%%%%%%%%%%%%%%%%%%%%%%%%%%%%%%%%%%%%%%

%https://tex.stackexchange.com/questions/168169/options-for-supplementary-materials-in-preprint-version-revtex-arxiv

%%%%%%%%%% Merge with supplemental materials %%%%%%%%%%
\pagebreak
\widetext
\begin{center}
\textbf{\large Supplemental Materials: 
Coherent Many-Body Oscillations Induced by a Superposition of Broken Symmetry States in the Wake of a Quantum Phase Transition
}
\end{center}
%%%%%%%%%% Merge with supplemental materials %%%%%%%%%%
%%%%%%%%%% Prefix a "S" to all equations, figures, tables and reset the counter %%%%%%%%%%
\setcounter{equation}{0}
\setcounter{figure}{0}
\setcounter{table}{0}
\setcounter{page}{1}
\makeatletter
\renewcommand{\theequation}{S\arabic{equation}}
\renewcommand{\thefigure}{S\arabic{figure}}
%\renewcommand{\bibnumfmt}[1]{[S#1]}
%\renewcommand{\citenumfont}[1]{S#1}
%%%%%%%%%% Prefix a "S" to all equations, figures, tables and reset the counter %%%%%%%%%%

This supplementary material outlines the complete solution of the quantum Ising chain that naturally leads to the new results presented in the main text. We believe the outline will be helpful to the reader by making our paper self-contained. The standard parts of the solution are based on three papers: the two older items~\cite{d2005,QKZteor-e} and the recent one~\cite{RadekNowak}. It begins from the very basics.

%%%%%%%%%%%%%%%%%%%%%%%%%%%%%%%%%%%%%%%%%%%%%%%%%%%%%%%%%%%%%%%%%%%%%%%%%%%%%%%%%%%
\section{ Quantum Ising chain }
\label{app:QIM}
%%%%%%%%%%%%%%%%%%%%%%%%%%%%%%%%%%%%%%%%%%%%%%%%%%%%%%%%%%%%%%%%%%%%%%%%%%%%%%%%%%%

The Hamiltonian for transverse field quantum Ising chain reads
\be
H~=~-\sum_{n=1}^N \left(  \sigma^z_n\sigma^z_{n+1}  + g\sigma^x_n\right)~,
\label{Hsigma}
\ee
where we consider a system of $N$ spins one-half with periodic boundary conditions, $ \vec\sigma_{N+1}~=~\vec\sigma_1$.  In the limit of $N\to\infty$, there are quantum critical points at $g_c=\pm1$, respectively, that separate the paramagnetic phase for $|g|>1$ from the ferromagnetic phase for $|g|$<1. For simplicity of presentation, we additionally assume that $N$ is even. The Jordan-Wigner transformation,
\bea
&&
\sigma^x_n~=~1-2 c^\dagger_n  c_n~, \\
&&
\sigma^z_n~=~
-\left( c_n + c_n^\dagger\right)
 \prod_{m<n}(1-2 c^\dagger_m c_m)~,
 \label{JW}
\eea
introduces fermionic annihilation ($c_n$) and creation ($c^\dagger_n$) operators. It maps the Hamiltonian in Eq.~\eqref{Hsigma} to
\be
 H~=~P^+~H^+~P^+~+~P^-~H^-~P^-~.
\label{Hc}
\ee
The projectors on subspaces with even ($+$) and odd ($-$) numbers of $c$-quasiparticles read
\be
P^{\pm}=
\frac12\left[1\pm\prod_{n=1}^N\sigma^x_n\right]=
\frac12\left[1~\pm~\prod_{n=1}^N\left(1-2c_n^\dagger c_n\right)\right].
\label{Ppm}
\ee
The reduced Hamiltonians in each parity subspace,
\be
H^{\pm}~=~
\sum_{n=1}^N
\left[
g \left(c_n^\dagger  c_n -\frac12\right) - c_n^\dagger  c_{n+1} + c_n  c_{n+1} \right]  +{\rm h.c.},
\label{Hpm}
\ee
differ in boundary conditions. Namely, in $H^-$ we assume periodic boundary conditions, $c_{N+1}=c_1$, and in $H^+$ we have antiperiodic boundary conditions, $c_{N+1}=-c_1$. 

The parity of the number of $c$-quasiparticles commutes with the Hamiltonian. As the ground state for  $g\gg1$ has even parity, we limit ourselves to that relevant subspace. The next step in the diagonalization of $H^+$ is a Fourier transform, 
\be
c_n~=~ 
\frac{e^{-i\pi/4}}{\sqrt{N}}
\sum_k c_k e^{ikn}~,
\label{Fourier}
\ee
with half-integer pseudo-momenta consistent with the antiperiodic boundary conditions,
\be
k~=~
\pm \frac12 \frac{2\pi}{N},
\pm \frac32 \frac{2\pi}{N},
\dots,
\pm \frac{N-1}{2} \frac{2\pi}{N}~.
\label{k}
\ee
After this transformation, the Hamiltonian takes the form
\bea
H^+~&=&
\sum_k
\left[
(g-\cos k) \left(c_k^\dagger c_k - c_{-k} c_{-k}^\dagger \right) 
+ \sin k
 \left(
 c^\dagger_k c^\dagger_{-k}+
 c_{-k} c_k
\right)
\right]~.
\label{Hck}
\eea
Its diagonalization is completed by a Bogoliubov transformation,
\be
c_k~=~
u_k  \gamma_k + v_{-k}^*  \gamma^\dagger_{-k}~,
\label{Bog}
\ee
where the Bogoliubov modes $(u_k,v_k)$ follow as eigenstates of the Bogoliubov-de Gennes equations,
\bea
\epsilon~ u_k &=& +2(g-\cos k) u_k+2\sin k~ v_k,\nonumber\\
\epsilon~ v_k &=& -2(g-\cos k) v_k+2\sin k~ u_k. \label{stBdG}
\eea
There are two eigenstates for each value of $k$, with eigenfrequencies $\epsilon=\pm\epsilon_k$, 
\be
\epsilon_k~=~2\sqrt{(g-\cos k)^2+\sin^2 k}~.
\label{epsilonk}
\ee
The eigenstates with positive frequency, $(u^+_k,v^+_k) = (\cos(\theta_k/2), \sin(\theta_k/2))$,
define a fermionic quasiparticle operator $\gamma_k~=~u_k^{+*} c_k + v^+_{-k} c_{-k}^\dagger$, where angles $\theta_k$ satisfy
$(\cos \theta_k, \sin \theta_k) = \frac{2}{\epsilon_k} (g-\cos k, \sin k)$. The negative frequency ones, with 
$(u^-_k,v^-_k)=(-v^+_k,u^+_k)$, formally define $\gamma_k^-=u_k^{-*}c_k+v_{-k}^-c_{-k}^\dagger=-\gamma_{-k}^\dagger$. 
After the Bogoliubov transformation, the Hamiltonian reads
\be
H^+~=~
\sum_k \epsilon_k~
\left( \gamma_k^\dagger \gamma_k~ - \frac12 \right).
\label{Hgamma}
\ee
Note that, due to the projector $P^+$ in Eq.~\eqref{Hc}, only states with even numbers of $c$-quasiparticles belong to the spectrum of $H$. 

The quasiparticle dispersion in Eq.~\eqref{epsilonk} implies a linear dispersion for small $k$ at the critical $g=1$, $\epsilon_k\approx 2|k|$, and the dynamical exponent $z$ is equal $1$. Moreover, for $k=0$, we have $\epsilon_0\propto |g-1|^1$ and
$z\nu=1$. Finally, the correlation length exponent $\nu$ is equal $1$. 

%%%%%%%%%%%%%%%%%%%%%%%%%%%%%%%%%%%%%%%%%%%%%%%%%%%%%%%%%%%%%%%%%%%%%
\section{ Linear quench and the Landau-Zener problem}
\label{sec:linearquench}
%%%%%%%%%%%%%%%%%%%%%%%%%%%%%%%%%%%%%%%%%%%%%%%%%%%%%%%%%%%%%%%%%%%%%

The Hamiltonian follows a linear ramp in the transverse field,
\be
g(t\leq0)~=~-\frac{t}{\tau_Q},
\label{linear}
\ee
with the quench rate $\tau_Q$. For convenience, here we fix the time when the ramp reaches $g(t_s)=0$ at $t_s=0$ (we use $t_s$ in the main text for clarity). As such, time $t$ runs from $-\infty$ to $0$ when the ramp stops at transverse field $g=0$, crossing the critical point at $g_c=1$ when $t_c=-\tau_Q$. The system starts in the ground state at $g\to\infty$, where $(u_k, v_k) = (1, 0)$, and is the vacuum state  annihilated by all corresponding Bogoliubov operators, $\gamma_k |0\rangle = 0$. 

In addressing the dynamical problem, it is convenient to employ the Heisenberg picture. The state of the system stays as the vacuum of Bogoliubov operators $\gamma_k$, while the Bogoliubov modes evolve according to the Heisenberg equation of motion $i\frac{d}{dt}c_k=[c_k,H^+]$. Following time-dependent Bogoliubov transformation,
\be
c_k = u_k(t)  \gamma_k + v_{-k}^*(t)  \gamma^\dagger_{-k},
\label{tildeBog}
\ee
this gives time-dependent Bogoliubov-de Gennes equations,
\bea
i\frac{d}{dt} u_k &=&
+2\left(g(t)-\cos k\right) u_k +
 2 \sin k~ v_k~,\nonumber\\
i\frac{d}{dt} v_k &=&
-2\left(g(t)-\cos k\right) v_k +
 2 \sin k~ u_k~,
\label{dynBdG} 
\eea
and the initial condition is $(u_k(-\infty),v_k(-\infty))=(1,0)$. 

Introducing a new time variable,
\be
\tau~=~4\tau_Q\sin k\left(\frac{t}{\tau_Q}+\cos k\right),
\label{tau}
\ee
that runs from $-\infty$ to $\tau^{\rm final}_k=2\tau_Q\sin(2k)$ for $t=0$,
allows one to map Eq.~\eqref{dynBdG} to the Landau-Zener (LZ) problem~\cite{d2005,QKZteor-e},
\bea
i\frac{d}{d\tau} u_k &=&
-\frac12 \tau\Delta_k~ u_k + \frac12 v_k,\nonumber\\
i\frac{d}{d\tau} v_k &=&
+\frac12 \tau\Delta_k~ v_k + \frac12 u_k.\label{LZ}
\label{BdGLZ}
\eea
Here, $\Delta_k=(4\tau_Q\sin^2 k)^{-1}$ sets an efficient rate of the transition for given $k$. 

Only modes with small $k$ that have small energy gaps at their anti-crossing point can get excited when the ramp is slow.
For such modes, $\tau^{\rm final}_k$ is much longer than the time when the anti-crossing is completed and we are allowed to use the LZ formula,
\be
p_k~\approx~
e^{-\frac{\pi}{2\Delta_k}}~\approx~
e^{-2\pi\tau_Qk^2}~,
\label{LZpk}
\ee 
where approximations become accurate for $\tau_Q\gg1$. 
Eq.~\eqref{LZpk} gives the probability that a pair of quasipartices with quasimomenta $+k$ and $-k$ got excited.
The mean density of kinks at $g=0$ is simply given by $n = \sum_k p_k / N$~\cite{d2005}. Taking the limit $N\to\infty$,
\be
n=\lim_{N\to\infty}\frac{1}{N}\sum_k p_k=
\frac{1}{2\pi}\int_{-\pi}^{\pi}dk~p_k\approx
\frac{1}{2\pi\sqrt{2\tau_Q}}.
\label{scaling}
\ee    
The density scales as an inverse of $\hat\xi\propto\tau_Q^{1/2}$, in full consistency with KZM prediction for $\nu=z=1$.
For convenience, we can use the density of kinks to supplement numerical prefactor
\be 
\hat\xi \equiv \frac{1}{n} = 2\pi\sqrt{2\tau_Q},
\label{hatxiprecise}
\ee 
making its inverse equal to mean density of kinks at the end of the ramp at $g=0$.

%%%%%%%%%%%%%%%%%%%%%%%%%%%%%%%%%%%%%%%%%%%%%%%%%%%%%%%%%%%%%%%%%%%%%%%%%%%%%%%%%%%%%%%%
\subsection{ Fermionic correlators }
\label{sec:Weber}
%%%%%%%%%%%%%%%%%%%%%%%%%%%%%%%%%%%%%%%%%%%%%%%%%%%%%%%%%%%%%%%%%%%%%%%%%%%%%%%%%%%%%%%%

To characterize the oscillations, we require more than just the excitation spectrum in Eq.~\eqref{LZpk}. 
A general solution to Eqs.~\eqref{BdGLZ} has the form~\cite{Damski_PRA_2006,QKZteor-e},
\bea
v_k(\tau)&=&
- a D_{-s-1}(-iz) - b D_{-s-1}(iz), \nonumber\\
u_k(\tau)&=&
\left(-\Delta_k\tau+2i\frac{\partial}{\partial\tau}\right)
v_k(\tau),
\label{general} 
\eea
where $D_m(x)$ is a Weber function, 
$s^{-1}=4i\Delta_k$, and $i z=\sqrt{\Delta_k}\tau e^{i\pi/4}$.
Constants $a,b$ are fixed by initial conditions. From the asymptotic behavior of the Weber function when $\tau\to-\infty$, one gets
$a=0$, and 
\be
|b|^2=\frac{e^{-\pi/8\Delta_k}}{4\Delta_k}~.
\ee 

We now focus at the end of the ramp when $t=0$ and $\tau=2\tau_Q \sin(2k)$. The argument of the Weber function now reads
$iz=2\sqrt{\tau_Q}e^{i\pi/4}{\rm sign}(k)\cos(k)$. Its absolute value is large for slow transitions (except near $k=\pm\frac{\pi}{2}$),
and one can again use the asymptotic behavior of the Weber function to find~\cite{QKZteor-e}:
\bea
|u_k|^2&=&
\frac12(1-\cos k)+p_k~,\nonumber\\
u_k v_k^* &=&
\frac12\sin k+
{\rm sign}(k)~
\sqrt{p_k(1-p_k)}~
e^{i\varphi_k}~,\nonumber\\
\varphi_k &=&
\frac{\pi}{4}+
2\tau_Q-
(2-\ln 4)k^2\tau_Q+
k^2 \tau_Q\ln\tau_Q-
\arg\left[\Gamma\left(1+ik^2\tau_Q\right)\right]~.
\label{att0smallk}
\eea
Above, $\varphi_k$ is a dynamical phase acquired by a pair of excited quasiparticles $(k, -k)$, with $\Gamma(x)$ being the gamma function. There are two combinations of $k$ and $\tau_Q$ appearing in those equations, which reflect two physical processes.
The first one is $\tau_Q k^2$, related to the KZM correlation length $\hat xi$. It determines the spectrum of excitations excited when crossing the critical point, $p_k$. The second is $k^2 \tau_Q\ln\tau_Q$, that implies a longer lenghtscale $\propto\sqrt{\tau_Q\ln\tau_Q}$, that appears due to the following adiabatic dephasing of excited quasiparticle modes, increasing of the dependence of dynamical phase $\varphi_k$ on $k$.

We follow~\cite{RadekNowak}, and approximate ${\rm arg} [\Gamma \left(1+i\tau_Qk^2\right)]\approx -\gamma_E \tau_Qk^2$ to make the phase more tractable. Here $\gamma_E$ is the Euler gamma constant. The approximation is valid for small enough $\tau_Qk^2$, which is consistent with the fact that excited quasiparticles have at most $\tau_Qk^2\approx 1/2\pi$, see Eq.~\eqref{LZpk}. This makes $\varphi_k$ conveniently quadratic in $k$,
\bea
\varphi_k-\varphi_0 = \left( \ln\tau_Q+\ln 4-2+\gamma_E \right) k^2\tau_Q 
                    = \left( \ln\tau_Q-0.036 \right) k^2\tau_Q 
            \approx k^2\tau_Q\ln\tau_Q.
            \label{quadraticvarphi}
\eea  
This makes it apparent that the dynamical phase is characterized solely by the second lengthscale $\propto\sqrt{\tau_Q\ln\tau_Q}$, that gives the scale on which the system has got effectively dephased.

The state of the system is Gaussian, and as such, it is fully determined by quadratic fermionic correlators. 
For our discussion here, the relevant one is the anomalous correlator,
\bea
\beta_R~\equiv~\langle c_{n+R} c_n \rangle = 
\frac{1}{\pi}\int_{0}^\pi dk~u_kv_k^*~\sin(kR).
\label{beta}
\eea
From Eq.~\eqref{att0smallk}, we get
\be 
\beta_R=\frac14 {\rm sign}(R)\delta_{|R|,1}+\delta\beta_R,
\label{betaR}
\ee 
The first term above is contributed by the ground state, and the second one is related to excitations,
\bea 
\delta\beta_R &=&
\frac{1}{\pi}\int_0^\pi dk~
\sqrt{p_k(1-p_k)}~
e^{i\varphi_k}
\sin(kR).
\label{deltabeta}
\eea
We approximate~\cite{RadekNowak},
\bea
\sqrt{p_k(1-p_k)} \approx
e^{-a\pi\tau_Q k^2}
A \sqrt{2 \pi} 
\left(\tau_Qk^2\right)^{1/2},
\label{A1}
\eea
making the integral analytically tractable. Above, $A$ and $a$ are the variational parameters, which can be optimally choosen as
$A\approx 19/20$ and $a\approx 4/3$. Putting together all those approximations we get,
\be 
\delta\beta_R\approx
\frac{e^{i\varphi_0}\sqrt{2}A}{\sqrt{\pi\tau_Q}} 
\int_0^{\infty} q
e^{-a\pi q^2+i q^2 \ln\tau_Q} 
\sin\frac{q R}{\sqrt{\tau_Q}} dq,
\label{deltabetaapp0}
\ee 
where $q=\sqrt{\tau_Q}k$ is a scaled pseudomomentum, and the upper limit of the integral have been safely extended to infinity, valid for slow transitions. Finally, this gives
\bea 
\delta\beta_R &\approx&
\frac{\sqrt{8\pi}A}{a^{3/2}}
\frac{R}{\sqrt{\hat\xi l^3}}
e^{ -\frac{2\pi}{a} (R/l)^2 }
e^{i\phi_R} 
=
C
n^2 R d~
e^{ -\frac{3\pi}{2} (R/l)^2 }
e^{i\phi_R}.
\label{deltabetaapp}
\eea 
Here, $C\approx57\sqrt{6\pi}/80$ is a constant, a phase factor
$\phi_R=
\frac14\pi+2\tau_Q-\frac32 {\rm arg}\left(1-\frac{3i\ln\tau_Q}{4\pi}\right)
-\frac98(R/l)^2\ln\tau_Q$, 
and the dephased correlation range is
\be 
l  = 
\hat\xi ~ \sqrt{ 1 + \left(\frac{3\ln\tau_Q}{4\pi}\right)^2 }.
\label{l}
\ee 
The range of this correlator becomes much longer than $\hat\xi$ for very slow quenches, when $\ln\tau_Q\gg4\pi/3$.
In addition to the factor $n^2\propto 1/\tau_Q$, its magnitude becomes significantly suppressed by a dephasing factor
\be 
d=\left(l/\hat\xi\right)^{-3/2}=
\left[ 1 + \left(\frac{3\ln\tau_Q}{4\pi}\right)^2 \right]^{-3/4}.
\label{d_lin}
\ee

%%%%%%%%%%%%%%%%%%%%%%%%%%%%%%%%%%%%%%%%%%%%%%%%%%%%%%%%%%%%%%%%%%%%%%%%%%%%%%%%%%%%%%%%
\section{ Transverse magnetization at the end of the linear ramp }
\label{sec:transend}
%%%%%%%%%%%%%%%%%%%%%%%%%%%%%%%%%%%%%%%%%%%%%%%%%%%%%%%%%%%%%%%%%%%%%%%%%%%%%%%%%%%%%%%%

At times later than $\hat t$, after crossing the critical point, the time-dependent Bogoliubov modes can be accurately decomposed as
\be 
\left(
\begin{array}{c}
u_k \\
v_k
\end{array}
\right)=
\sqrt{1-p_k}
\left(
\begin{array}{c}
u^+_k \\
v^+_k
\end{array}
\right)+
\sqrt{p_k}
\left(
\begin{array}{c}
v^+_k \\
-u^+_k
\end{array}
\right)
e^{i\varphi_k},
\label{uvafterhatt}
\ee
were $(u^+_k,v^+_k)$ is a positive-frequency stationary Bogoliubov mode at $g(t)$. The transverse magnetization is
\bea 
\langle \sigma^x \rangle 
=
\frac{1}{2\pi} \int_{-\pi}^{\pi}dk~\left(|u_k|^2-|v_k|^2\right)
=
\langle \sigma^x \rangle_{\rm GS}
-\frac{2}{\pi} \int_0^{\pi} dk \left({u^+_k}^2-{v^+_k}^2\right) p_k  + 
\frac{4}{\pi} \int_0^{\pi} dk~ u^+_k v^+_k \sqrt{p_k(1-p_k)}  \cos\varphi_k.
\eea
Here $\langle \sigma^x \rangle_{\rm GS}=\frac{1}{\pi} \int_{0}^{\pi}dk~\left({u^+_k}^2-{v^+_k}^2\right)$ is a stationary transverse magnetization in the instantaneous ground state at $g(t)$.

Focusing on the final $g=0$, we have $u^+_k=\sin(k/2)$, $v^+_k=\cos(k/2)$, and $\langle \sigma^x \rangle_{\rm GS}=0$.
For the small $k$ in the support of $p_k$, we can approximate ${u^+_k}^2-{v^+_k}^2\approx -1$ and $u^+_k v^+_k \approx k/2$. Consequently, 
\bea 
\langle \sigma^x \rangle 
=
\frac{2}{\pi} \int_0^{\pi} dk~ p_k  + \frac{2}{\pi} \int_0^{\pi} kdk~ \sqrt{p_k(1-p_k)}  \cos\varphi_k 
\approx
2n + 2 {\rm Re}~ \delta\beta_1 
\approx
2n + 2C d~ n^2 \cos\phi_0,
\label{sigmaxend}
\eea
where in the last step we assumed that $R=1 \ll l$ in Eq.~\eqref{deltabetaapp}. 

%%%%%%%%%%%%%%%%%%%%%%%%%%%%%%%%%%%%%%%%%%%%%%%%%%%%%%%%%%%%%%%%%%%%%%%%%%%%%%%%%%%%%%%%
\section{ Transverse magnetization after the linear ramp }
\label{sec:transosc}
%%%%%%%%%%%%%%%%%%%%%%%%%%%%%%%%%%%%%%%%%%%%%%%%%%%%%%%%%%%%%%%%%%%%%%%%%%%%%%%%%%%%%%%%

Waiting for time $t$ after the ramp stops at $g=0$, where all quasiparticle frequencies become degenerate with $\epsilon_k=2$, the dynamical phase $\phi_0$ in Eq.~\eqref{sigmaxend} evolves into $\phi_0+4t$, and the transverse magnetization oscillates as
\bea 
\langle \sigma^x(t>0) \rangle_{KZM} = 2n + 2C d~ n^2 \cos(4t+\phi_0).
\label{sigmaxKZ}
\eea
This is a partial contribution to the transverse magnetization, that comes only from the KZM excitations localized near $k=0$. It turns out, that it has to be supplemented with a wide spectrum of excitations induced by the sharp end of the ramp at $g=0$, where the time derivative of $g(t)$ is discontinuous.

For modes beyond the support of KZM excitation spectrum, this small excitation can be calculated with the help of the adiabatic perturbation theory. Each Landau-Zener system follows the adiabatic positive-frequency mode, plus a small excitation amplitude $B_k$ for the negative-frequency one:
\be 
\left(
\begin{array}{c}
u_k \\
v_k
\end{array}
\right)\approx 
\left(
\begin{array}{c}
u^+_k \\
v^+_k
\end{array}
\right)+
B_k
\left(
\begin{array}{c}
v^+_k \\
-u^+_k
\end{array}
\right).
\ee
For the time dependence ending with a discontinuous time derivative at time $t_0$, the leading contribution from the final discontinuity to this small excitation amplitude is 
\be 
B_k(t_0)\approx \int^{t_0} dt~ e^{ 2i \epsilon_k[g(t_0)] (t_0-t) } \dot\Theta_k(t).
\ee 
Here $\dot\Theta_k=u^+_k \frac{d}{dt} v^+_k - v^+_k \frac{d}{dt} u^+_k$. A suitable integration by parts yields
\be 
B_k \approx i \frac{\dot\Theta_k(t_0)}{2\epsilon_k[g(t_0)]},
\ee 
where we neglected higher order derivative terms, that are also of higher order in $1/\tau_Q$.
For $g(t_0)=0$, we have $\epsilon_k(0)=2$. 
From the stationary Bogoliubov modes, we obtain a formula for $\dot\Theta_k(t_0)$ and
\be 
B_k(0) = i\frac{\sin k}{8\tau_Q}.
\ee
Waiting at $g=0$ for time $t$ after the end of the linear ramp, the amplitude picks up a extra dynamical phase factor,
\be 
B_k(t>0) = i\frac{\sin k}{8\tau_Q} e^{4it},
\ee 
leading to
\bea
|u_k|^2 &=&
{u^+_k}^2 -  u^+_k v^+_k \frac{\sin k}{4\tau_Q} \sin 4t + {\cal O}\left(\tau_Q^{-2}\right).
\eea
Here, ${u^+_k}^2$ is the ground state contribution that has already been taken into account in Eq.~\eqref{sigmaxKZ}.
With $u^+_k v^+_k = \frac12 \sin k$, the extra contribution to the transverse magnetization excited by the abrupt termination of the linear ramp becomes 
\be 
\langle \sigma^x(t>0) \rangle_{\rm disc} = -\pi^2n^2\sin(4t).
\ee
Assembling it together with the KZM part in~\eqref{sigmaxKZ}, a total transverse magnetization after the end of the linear ramps is
\be 
\langle \sigma^x (t>0) \rangle = 
2n + 2C d~ n^2 \cos(4t+\phi_0) - \pi^2 n^2 \sin(4t).
\ee 
Alternatively, we can write it as
\begin{eqnarray}
    \label{1d_Xt}
    \langle \sigma^x(t>0) \rangle &=& 2n + A \cos(4t+\tilde{\phi}_0).   
\end{eqnarray}
Here, the amplitude and the phase satisfy
\bea
    A &=& n^2 \sqrt{\pi^4+4\pi^2 C d \sin\tilde\phi_0 +4C^2d^2} \\ \nonumber
    \tan\tilde\phi_0&=&\tan\phi_0+\frac{\pi^2}{2Cd\cos\phi_0}.
    \label{1d_amplitude}
\eea
Notice that the amplitude depends on $\tau_Q$ not only through the power law $n^2\propto\tau_Q^{-1}$, but also through a further non-universal modulation. 

%%%%%%%%%%%%%%%%%%%%%%%%%%%%%%%%%%%%%%%%%%%%%%%%%%%%%%%%%%%%%%%%%%%%%%%%%%%%%%%%%%%%%%%%
\section{ Oscillations after the end of a smooth ramp }
\label{sec:transoscsmooth}
%%%%%%%%%%%%%%%%%%%%%%%%%%%%%%%%%%%%%%%%%%%%%%%%%%%%%%%%%%%%%%%%%%%%%%%%%%%%%%%%%%%%%%%%

Smoothing out the final time-derivative discontinuity removes $\langle \sigma^x \rangle_{\rm disc}$, in the sense that it makes it of higher order in $n$ in comparison to the leading KZM contribution $\propto n^2$. It does not affect $p_k$, as long as the linearity of the ramp within $\pm\hat t$ of the critical point is not altered. However, the longer evolution between $g=1$ and $g=0$ does affect dynamical phase $\varphi_k$, adding time for extra dephasing. 

For $k$ within the support of $p_k$, and $g$ far enough from the critical $1$, we can approximate the quasiparticle spectrum as
\be 
\epsilon_k \approx 2|1-g| + \frac{g}{|1-g|} k^2.
\label{epsilonkapprox}
\ee 
Now, we can compare the dynamical phase accumulated during the linear ramp and the smooth one. Both ramps are described by 
\be 
g_j(t)=1-\epsilon_j(t),
\ee 
For the familiar linear ramp,
\be 
\epsilon_l(t)=\frac{t-t_c}{\tau_Q},
\ee 
where $t' = t-t_c$ is the time measured with respect to the moment when the critical point is crossed. For linear evolution $t'$ extends to $\tau_Q$. 
As an example of the smooth ramp we choose,
\be 
\epsilon_s(t)=\frac{t-t_c}{\tau_Q} - \frac{4}{27}\left(\frac{t-t_c}{\tau_Q}\right)^3.
\ee
Here $t'$ extends to $3\tau_Q/2$. At the final $g=0$, the dynamical phases for the smooth and the linear ramps differ by
\bea 
\delta\varphi_k &=& 
2 \int_0^{3\tau_Q/2} dt'
\left[ 2|1-g_s(t')| + \frac{g_s(t')}{|1-g_s(t')|} k^2 \right] 
-
2 \int_0^{\tau_Q} dt'
\left[ 2|1-g_l(t')| + \frac{g_l(t')}{|1-g_l(t')|} k^2 \right].
\eea 
Performing the integration, we obtain
$ 
\delta\varphi_k=\frac{7}{16}\tau_Q+0.2164~\tau_Q k^2.
$ 
Accordingly, Eq. \eqref{quadraticvarphi} is modified to 
$
\varphi_k-\varphi_0=k^2\tau_Q\ln \tau_Q+0.2164~k^2\tau_Q.
$ 
Therefore, a simple replacement of $\ln\tau_Q$ in~Eq.~\eqref{d_lin} with $\ln\tau_Q+0.2164$ yields a new dephasing coefficient,
\begin{equation}
\tilde d =
\left[ 
1 +
\left( 
\frac{3(0.2164+\ln\tau_Q)}{4\pi} 
\right)^2
\right]^{-3/4}.
\label{d_smooth}
\end{equation}
This is a minor change for $\tau_Q\gg1$. A new phase becomes
\begin{equation}
    \tilde\phi_0=\frac{7}{16}\tau_Q+\frac14\pi+2\tau_Q-\frac32 {\rm arg}\left(1-i\frac{3\ln\tau_Q}{4\pi}\right).
\end{equation}
Here, the change is $\propto\tau_Q$. Finally, the transverse magnetization after the end of the smooth ramp oscillates as
\be 
\langle \sigma^x(t>t_s) \rangle = 
2n + 2C \tilde d n^2 \cos\left(4(t-t_s)+\tilde\phi_0\right),
\ee 
where $t-t_s$ is the free evolution time. The amplitude of the transverse oscillations simply depends on $n^2\propto\tau_Q^{-1}$ up to a logarithmic correction introduced by the dephasing factor $\tilde d$, with no further modulation of the amplitude due to interference between the KZM part and the part due to discontinuity at the end of the ramp.

%%%%%%%%%%%%%%%%%%%%%%%%%%%%%%%%%%%%%%%%%%%%%%%%%%%%%%%%%%%%%%%%%%%%%%%%%%%%%%%%%%%%%%%%
\section{ Oscillations during the ramp }
\label{sec:transoscduring}
%%%%%%%%%%%%%%%%%%%%%%%%%%%%%%%%%%%%%%%%%%%%%%%%%%%%%%%%%%%%%%%%%%%%%%%%%%%%%%%%%%%%%%%%

The expansion in Eq.~\eqref{uvafterhatt} is accurate at any time later than $\hat t$ after the phase transition. 
The phase $\varphi_k$ increases as
\be 
\varphi_k(t)=\int^t dt'~2\epsilon_k[g(t')].
\ee 
After $\hat t$, the quasiparticle spectrum for KZM excitations, that are localized near $k=0$, can be considered flat and equal to the gap $\epsilon_0(g)$ that opens with the increasing distance from the critical point. Therefore, the transverse field oscillates with frequency given by twice the instantaneous gap. 

Beyond the approximation of flat dispersion, there is some dephasing. The dephasing time can be estimated with the help of the approximate dispersion relation in Eq.~\eqref{epsilonkapprox}. The difference between $\epsilon_k$ for $\hat k\approx1/\sqrt{\tau_Q}$ and $k=0$ is $\delta\epsilon \approx \frac{g}{|1-g|\tau_Q}$. Therefore, the phase gets scrambled on a timescale
\be 
\tau_D = \frac{\pi}{\delta\epsilon} \approx \frac{\pi|1-g|\tau_Q}{g}.
\ee 

%%%%%%%%%%%%%%%%%%%%%%%%%%%%%%%%%%%%%%%%%%%%%%%%%%%%%%%%%%%%%%%%%%%%%%%%%%%%%%%%%%%%%%%%
\section{ Ramp termination at finite transverse field }
\label{sec:transoscfinite}
%%%%%%%%%%%%%%%%%%%%%%%%%%%%%%%%%%%%%%%%%%%%%%%%%%%%%%%%%%%%%%%%%%%%%%%%%%%%%%%%%%%%%%%%
In case the linear ramp is stopped at a finite $g_f$, but after $+\hat t$, the quasiparticles continue to evolve with a dispersion $\epsilon_k=2[(\cos k-g_f)^2+\sin^2k]^{1/2}$. The non-KZM excitations dephase after time $\propto (\epsilon_\pi-\epsilon_0)^{-1}=1/4g_f$. The KZM excitations, up to $\hat k^2=1/2\pi\tau_Q$, dephase after time $\propto(\epsilon_{\hat k}-\epsilon_0)^{-1}\approx2\pi\tau_Q (1-g_f)/g_f$. It is much longer than the non-KZM dephasing time provided that $\tau_Q(1-g_f)\gg1/8\pi$. The frequency of the KZM oscillations is set by twice the quasiparticle gap, $2\epsilon_0=4(1-g_f)$. They are much faster than their dephasing, provided that $g_f\ll 8\pi\tau_Q(1-g_f)^2$. With the other condition, $\tau_Q(1-g_f)\gg1/8\pi$, we arrive at $g_f\ll 1-g_f$, or, equivalently, $g_f\ll1$ as a necessary condition for pure KZM oscillations that are much faster than their dephasing.
%

%%%%%%%%%%%%%%%%%%%%%%%%%%%%%%%%%%%%%%%%%%%%%%%%%%%%%%%%%%%%%%%%%%%%%%%%%%%%%%%%%%%%%%%%
\section{Comments on the numerical simulations}
%%%%%%%%%%%%%%%%%%%%%%%%%%%%%%%%%%%%%%%%%%%%%%%%%%%%%%%%%%%%%%%%%%%%%%%%%%%%%%%%%%%%%%%%
The behavior of the transverse magnetization during the quench in the two-dimensional quantum Ising model, presented in Fig.~3 of the main text, have been obtained with the help of numerical simulations based on tensor networks.

To simulate the time-evolution in a finite lattice of $11 \times 11$ spins, we employ the time-dependent variational principle for matrix product states (MPS) algorithm~\cite{Haegeman2016}. We use 4th order time-dependent Suzuki-Trotter decomposition, and MPS bond dimensions up to $384$ to verify convergence. We focus on magnetization in the center of the lattice. To increase the numerical stability, we placed the spins neighboring the center next to each other in a one-dimensional MPS chain that spans the two-dimensional lattice. We additionally merged groups of $5$ spins, e.g., the central spin and its four neighbors. Such enlarged local sites of MPS come at an increased numerical cost, but it helps to increase numerical stability.

Simulation in the thermodynamic limit has been performed using two-dimensional iPEPS ansatz with the time-evolution based on the neighborhood tensor update algorithm of Ref.~\cite{ntu,dziarmaga2021simulation}. The presented data have been obtained for iPEPS bond dimensions up to $8$ and timestep of $0.01$ in a 2nd order Suzuki-Trotter decomposition. 

%%%%%%%%%%%%%%%%%%%%%%%%%%%%%%%%%%%%%%%%%%%%%%%%%%%%%%%%%%%%%%%%%%%%%%%%%%%%%%%%%%%%%%%%%%%%%%%%%%%%%%%%%%%%%%
\end{document}